\RequirePackage{fix-cm}
\documentclass[]{elsarticle}
\usepackage{geometry}
\usepackage{setspace}
\usepackage{bm}
\usepackage{tikz,pgfplots}
\usepackage{graphicx} 
\usepackage{etoolbox}
\apptocmd{\sloppy}{\hbadness 10000\relax}{}{}
\usepackage[rightcaption]{sidecap}
\usepackage{microtype}
\usepackage{subfig}
\usepackage{wrapfig}
\usepackage{ragged2e}
\usepackage{color}
\usepackage{amsmath}
\usepackage{color}
\usepackage{setspace}
\begin{document}

\begin{frontmatter}

\title{Compromised account detection using authorship verification: a novel approach}


\author{Forough Farazmanesh\fnref{myfootnote2}}
\fntext[myfootnote2]{Email:ffarazmanesh@gmail.com}

\author{Fateme Foroutan\fnref{myfootnote3}}
\fntext[myfootnote3]{Email:Foroutan@gmail.com}

\author{Amir Jalaly Bidgoly \fnref{myfootnote1}}
\fntext[myfootnote1]{Corresponding author;Assistant Professor; Email:Jalaly@qom.ac.ir}


\address{Department of Computer Engineering, University of Qom, Qom, Iran}

\begin{abstract}
Compromising legitimate accounts is a way of disseminating malicious content to a large user base in Online Social Networks (OSNs). Since the accounts cause lots of damages to the user and consequently to other users on OSNs, early detection is very important. This paper proposes a novel approach based on authorship verification to identify compromised twitter accounts. As the approach only uses the features extracted from the last user's post, it helps to early detection to control the damage. As a result, the malicious message without a user profile can be detected with satisfying accuracy. Experiments were constructed using a real-world dataset of compromised accounts on Twitter. The result showed that the model is suitable for detection due to achieving an accuracy of 89\%.

\end{abstract}
\begin{keyword}
\texttt  Compromised accounts  \sep Hacked accounts  \sep Twitter \sep Authorship verification  \sep LSTM \sep Natural language processing
\end{keyword}

\end{frontmatter}
\section{Introduction}
\indent Online Social Networks (OSNs) allow people to easily share their thoughts and ideas about any subject with others. OSNs accounts are valuable for cybercriminals to further his own malicious goals since those are having access to the network of trusted friends that controls how information spreads in the social network. Because OSNs users are more likely to respond to messages coming from friends, a cybercriminal can exploit this trust by comprising accounts to speed up the spread of malicious content \cite{jagatic2007social}.

\indent A compromised account is a legitimate account that has been hijacked by a hacker for distributing fake and harmful content \citep{igawa2015recognition}. Since the accounts must be returned to their respective owners after they are detected, those can be not removed \citep{egele2013compa}. attackers take control of a popular account to send their message or disseminate fake information to a large user base. The compromised accounts cause various kinds of damage, like a tarnished reputation and financial losses, to the user and subsequently to other users on the network \cite{egele2015towards}.

\indent The fact 
that about 160000 Facebook accounts are compromised every day shows that compromising legitimate accounts is a serious threat to OSNs security. So an important task among security researchers is the detection of compromised accounts. In this direction, researchers proposed several approaches using various measures usually associated with data gathered from the user in the network. The approaches include domains such as statistical analysis, behavioral analysis, text analysis, and machine learning.

\indent Most of the previous works were focused on the profiling of user behavior patterns using different features to detect unusual deviations from one's behavior. To extract the usage pattern, data such as the text content, the time print, and click stream are analyzed. But the dynamic behavior of the legitimate user of an OSNs account makes these approaches challenging.

\indent This work introduces the approach that can reliably detect compromised accounts. The approach
focuses on user Authorship verification checking whether an author's linguistic style of new post matches another author's linguistic style \citep{brocardo2013authorship, rocha2016authorship}. In other words, each new post is examined whether the post is possible to match to posts written by the legitimate user. In summary, this paper makes the following contributions:

\begin{enumerate}
\item This paper introduce the idea of using authorship verification based on deep learning techniques to detect compromised Twitter accounts.
\item We have shown that one can achieve acceptable accuracy having only one post per user without having a user profile. The benefits of this work are its high speed and usability in low-profile accounts.
\item Experiments were performed using a real dataset consisting of tweets, which are considerably shorter than those like email conversations, or online documents. Using short text messages means a greater challenge, but it also provides an opportunity to address the potential of the novel approach. the results also showed that despite the limited lengths of the posts, it is possible to detect comprised account reliability using the well-established approach in natural language processing.
\end{enumerate}

\indent The remaining of the work is organized as follows: section 2 presents an overview towards compromised accounts. In section 3, we formally define the compromised account detection problem. Section 4 states details about the proposed model. Section 5 evaluate the model along dataset used in the experiments. Section 6 states our conclusions.

\section{Related Work}
Researchers have already shown that finding users' passwords with attacks like  dictionary attack is easier than expected \cite{zipflaw}.
In recent years various ways such as statistical analysis, text mining, machine learning-based, behavioral profile-based, and hybrid approaches have been employed by researchers to identify compromised accounts. In the approaches available, a variety of OSNs data like post content, time print, and user profile has been used. Using this data the behavior and the usage pattern actual owner of the related account is extracted. But this method has been challenged due to the need for high data as well as the dynamic behavior of the legitimate user.

Some researchers have analyzed text messages to identify compromised accounts. Researchers in \cite{igawa2015recognition,igawa2016recognition} have considered only text-based data and examined authorship verification using N-grams extracted from user posts. The accuracy of the method was in the range of 73\% to 95\% on a sub-sample of news articles and blog posts collected from 1 million online sources. The method does not work well for short texts because the writing style of extracting short texts may be the same for many users.
Recent research is extracted user's writing style based on his previous posts to create a baseline. To detect a hacked account, the baseline will be matched with the user's next posts and baseline updating will be done to keep the model up to date . They achieved 93\% accuracy using a dataset of 1000 user profiles \cite{barbon2017authorship}. This approach detects the change in writing styles of high-profile and active accounts only. If the account is less frequently used, the false positive will be high.

Recent work has divided the accounts into two categories, namely user and spam, with the statistical text analysis and KL-Divergence method \cite{seyler2018identifying}. Experiments were conducted on 467 million of the twitter corpus. This approach achieved 80\% accuracy in detecting compromised accounts. In this analysis, word similarity is not considered and the language model used is simple.

Statistical modeling with anomaly detection is one of the approaches to detect compromised OSNs accounts. Various features like a time-stamp, location, direct user interaction, proximity, message source, and message topic have used by some researchers to identify whether an account is compromised or not. In this approach anomalous behavior is diagnosed by comparing normal model \cite{seyler2018identifying,egele2015towards}.  In \cite{viswanath2014towards}, the Yelp dataset consists of 92725 Yelp reviewers and the Twitter dataset consists of a random sample of 100K out of the 19M Twitter users is used to perform this analysis. Detection rate over 66\% is achieved using unsupervised anomaly detection techniques, Principal Component Analysis (PCA). The challenges of these approaches include: the need for high profile accounts for proper detection and decreased accuracy due to dynamic user behavior. 

Egele et al. provided a tool called COMPA combining introversive and extroversive behavior and achieved 98.6\% accuracy to detect compromised accounts. Features such as first activity, activity preference, activity sequence, and action latency are introversive. Feature such as browsing preference, visit duration, request latency, and browsing sequence are introversive. 50 facebook users recruited for gathering experiments data \cite{ruan2015profiling}. This approach works poorly for accounts with low-profile. Also since no clickstream behavior is available for automated attacks based on APIs, intentions of this approach cannot be applied in practice.

Reviewing existing research in this field can be concluded that the challenges in this area include:
\begin{enumerate}
\item The dynamic behavior of the legitimate user of an account has other challenged the detection. 
\item It is difficult to identify Compromised accounts with low-profiled.
\item Since as a solution, you cannot block a Compromised account, due to its bad impact to the legitimate user, speed detection is an advantage.
\end{enumerate}
We already shows that considering the text, we can identify the fake news \cite{fakenews}. In a similar way in this paper by covering these challenges, we show that using only a single user post without having a user profile can distinguish authorship of social network posts with acceptable accuracy.

\section{PROBLEM STATEMENT}
In this section, we introduce some mathematical notations and formally define the problem.

\textbf{Compromised} \textbf{account}. 
A compromised account is one accessed by a person, like Criminals and hackers, not authorized to use the account. 
A normal post is when the authorized user publishes the post. 
At some point in time, i.e. compromised point, an unauthorized user will gain control and publishes so called compromised posts.

\textbf{Healthy} \textbf{account}. A standardize healthy account is one that only the legitimate user has access to.

\textbf{Notations}. Let $A=\{a_1, a_2, a_3,..., a_n\}$  denote a dataset of $n$ twitter accounts which we crawl to evaluate our model. Each twitter accounts $a_i$ is associated with a set of posts denoted as $P_i=\{p_{i_1}, p_{i_2}, p_{i_3}, ..., p_{i_m}\}$. Additionally, let $T= \{t_1, t_2, t_3, ... , t_m\}$ denote a set of time points Which shows the accounts' compromised point as follow:
If the account $a_i$ is healthy then $t_i=\infty$.
If the account $a_i$ is comprised then $t_i=k$ show compromised point(Time indicates the number of posts.).

Using the symbols and definitions mentioned above, the problem of compromised account detection are defined as follows:

For each pair of posts $(p_{i_k}, p_{j_r})$, we aim to learn the model $M$, which can predict whether the user of the publisher of both posts is the same, in other words, is $i$ equal to $j$. 

The model $M$ is a classifier that can distinguish compromised from normal user accounts based on a set of features derived from LSTM. This approach based on the trained model $M$ for every new post have to decide whether the account is malicious or normal.

\section{PROPOSED MODEL}

In this work, we present an approach for compromised account detection based on authorship verification. Our method can recognize the compromised account by analyzing user text posts.  This way does automatic feature extraction by LSTM as an account writing style. It converts each post into feature vectors then learns what differentiates are between the posts features of two different authors. According to the differences, it detects whether an account has been compromised.

The proposed approach consists of two major phase: training phase and application phase. Training phase consists of three major steps: data augmentation, feature extraction, learn binary classifier.

\begin{figure}[h!]
\centering
\includegraphics[scale=0.75]{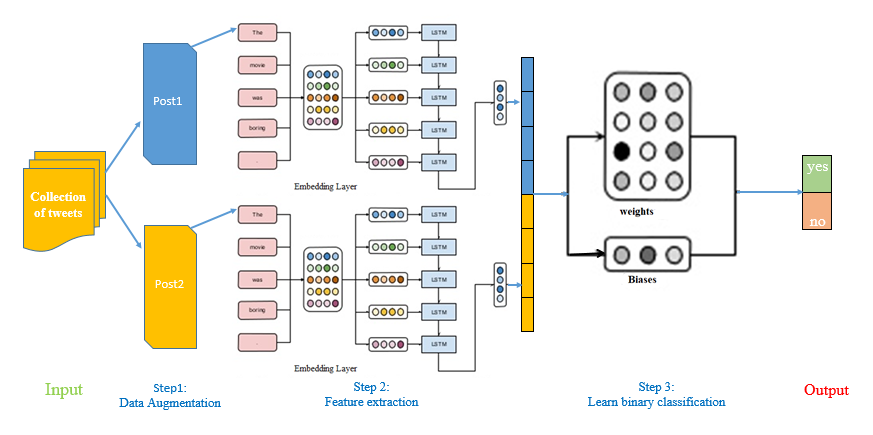}
\caption{Training phase in proposed approach.}
\label{fig:model1}
\end{figure}

\begin{figure}[h!]
\centering
\includegraphics[scale=0.75]{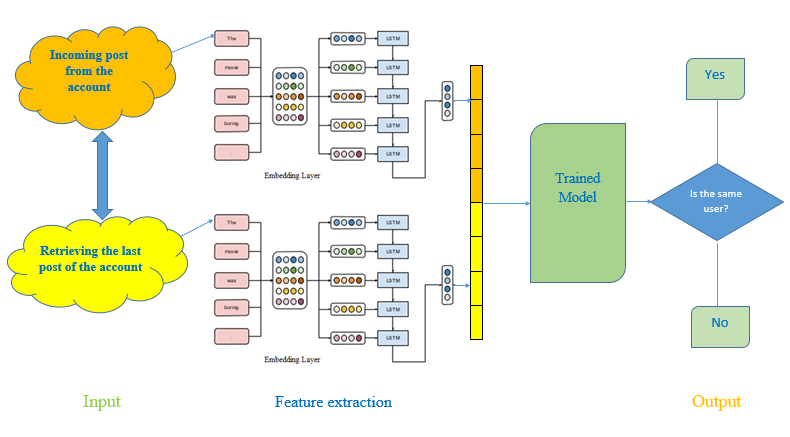}
\caption{Application phase in proposed approach.}
\label{fig:model2}
\end{figure}

The approach works as follows: In the training phase as shown in figure \ref{fig:model1}, a model is trained to learn the extent of the permissible difference in writing style between a user's posts to not flag them as malicious. Finally, based on these differences, it determines whether the author of the new post is the legitimate user of the account or not.  In the application phase as shown in figure \ref{fig:model2}, for each new post from the account, the last post that the user has authored is retrieved. Then new representations are separately made of both posts (new post and the last post). Using the representations as inputs, the model can determine whether the new post is malicious or not.

In the following, we describe training phase in more detail. 

\textbf{A. Data augmentation.} In order to improve model generalization, The raw data was converted to augmented data. In this process the raw data which includes $UserID$ (include $a_i$ to $a_n$), $post$ (include $p_i$ to $p_n$) and tag (comprised or not comprised) is converted to augmented data including $UserID_1$ and $post_1$ and $UserID_2$ and $post_2$ and label.

Data augmentation is done according to equations 1 and 2 as follow:

\begin{equation}
TrainData=\{(p_{ik},p_{jr})| p_{ik} \in p_i , p_{jr} \in p_j , k< t_i \quad  \textrm{and} \quad  r< t_j \} \quad
\centering Lable\big((p_{ik},p_{jr})\big)=  \left\{
\begin{array}{ll}
      1 & i=j \\
      0 & i\ne j \\
\end{array} 
\right. 
\end{equation}

\begin{equation}
TestData=\{(p_{ik},p_{ir})| t_i\ne\infty \Leftrightarrow  (k< t_i \quad  \textrm{and} \quad  r\geq t_i) \} \quad
\centering Lable\big((p_{ik},p_{ir})\big)=  \left\{
\begin{array}{ll}
      1 & t_i\ne\infty \\
      0 & t_i=\infty \\
\end{array} 
\right. 
\end{equation}

\textbf{B. Feature extraction}. The approach to the problem is to extract the linguistic styles of tweets (Twitter messages), by converting them into feature vectors. Feature extraction is done in two steps:

\textbf{B.1. Embedding words}. First, words from the vocabulary are mapped to vectors using Glove and Word2vec. The purpose is comparing these two embeddings to discover which one produces better results.
In embedding words, each word as a d-dimensional vector is represent regarding the relationships that may exist between the words in a context. Word2Vec \cite{mikolov2013distributed} learn word embeddings using shallow neural network while Glove  \cite{pennington2014glove} produce them based on co-occurrence statistics from a corpus.

\textbf{B.2. Extract user writing style by LSTM}. In this step, LSTM used to generate feature representations from tweets.

 LSTM as a  special version of Recurrent Neural Network (RNN). RNN is a generalization of feedforward neural network, which has a recursive nature as it executes an the same function for each input and the output of each step is used as the next step input. The distinctive feature of LSTM is that it remembers the patterns in the data for a longer period of time. LSTM has been widely used for many tasks where cases  are successively correlated \cite{bahdanau2014neural,ren2016deceptive}. For example, LSTM can be used in applications such as unsegmented, connected handwriting recognition \cite{graves2008novel} or speech recognition \cite{sak2014long, li2015constructing}.  In the LSTM, the information flow is controlled through three types of gates: input gate, output gate, and forget gate. Things are selectively forgotten or remembered by LSTM.
 
to extract user writing style, Two LSTM is concurrently trained on the augmented training set.  Each LSTM is used to learn a representation of the user's post as their writing style. Accordingly,for all training data, at each time step t for both posts(identical or different) i and j, The first LSTM takes the $i-th post$ embedding vector \(e_k\) as input and The second LSTM takes $j-th post$ embedding vector as input and they learn the representation of $post i and j$.

\textbf{C. learn binary classification}. After extracting features, those be used as input a neural network model used as a classifier. We turn the compromised account detection task into a binary classification problem. The goal of the classifier is to decide for each new post whether it is malicious or not. this binary classification learn what differentiates two different authors within a
corpus goal can be formulated as to successfully identify whether the authors of a pair of documents are identical.

The this step comprehends binary classification to compare the feature posts and check whether the authors of a pair of posts are identical.
 In the experiment, An simple network was used as classifier, implemented with the Tensor Flow6 library.

\section{EVALUATION }

To answer how effectively the proposed model detects compromised account, we evaluated it on real dataset from Twitter. In the following we first detail the dataset used to evaluate  performance of the approach. Subsequently, we discuss the training and testing phase.
 
\textbf{A. Data collection and annotation.} As there is no dataset publicly available for this work, we crawl a real dataset containing compromised accounts. This dataset contains tweets collected from Twitter in which the messages are limited on 140-character length. Using short text messages in classification makes a greater challenge, but it also provides an opportunity to address the power of this approach.
Our dataset includes two types of accounts: compromised account and healthy account. Also, a compromised account has two types of tweets: tweets written by a hacker, as compromised tweets, and tweets from the original user, as normal tweets. 

The first 10 normal accounts and 10 compromised accounts are identified. Compromised accounts were those self-reported being hacked. The accounts' tweets were collected using the Twitter Streaming API. The dataset is composed of 20 user and over 86347 tweets collected during January 2018. To specify whether tweets type is compromised or normal, these tweets are manually annotated. 

\textbf{B. Data pre-processing.} VanDam et al. consider patterns of the compromised tweets compared to the normal tweets to identify importance of meta information for detecting hacked accounts. The meta information examined are hashtags, mentions and URLs used in tweets, source used to post the tweet, sentiment (positive or negative), and retweets. They found, in addition to the tweet content, this patterns can improve performance of compromised tweet detection. Also they reported terms and sources were the best data to predict the hacked accounts\cite{vandam2017understanding}. As the content of user posts can be a useful source for detecting hacked accounts, user posts along with its metadata, such as hashtags and URLs, were used in this paper.

\textbf{D. Evaluation metrics.}  Following metrics are used to evaluate results:

\begin{itemize}
\item{\textbf{Accuracy:} Accuracy is ratio of the number of the tweets correctly classified to the total number of tweets under consideration.}
\item{\textbf{F-measure:} F-measure is is the harmonic mean of the precision and recall. Precision is defined as proportion of hacked accounts identifications was actually hacked. Recall is defined as proportion of actual hacked accounts was identified correctly.}
\end{itemize}

\textbf{E. Training phase.} In this section, we answer the following questions:
\begin{enumerate}
\item{it possible to detect malicious messages using the suggested approach with the short length of social network posts?} 

\item{As a subproblem, we evaluate what embedding words, i.e. glove or word2vec, are appropriate to generate posts representation to identify malicious posts in the context of very short texts (at most 140 characters).} 
\end{enumerate}

For this purpose we train on the 80\% real dataset two models, one with Glove embedding and other with Word2vec.

\textbf{F. Testing phase and results.} To evaluate the approach efficiency, standard metrics  i.e. accuracy and F-measure are used. The results are summarized in Table \ref{table:1}. Results show that the novel approach can detect comprised account with up to $87\%$ accuracy. When model is trained with Glove embedding reaches an accuracy of $89\%$ and when  model is trained with Word2vec embedding reaches an accuracy of $87\%$. As a result, using short text without any additional information this approach can accurately detect malicious messages. Given the acceptable accuracy, the following advantages can be derived for this approach:
  
\begin{table}[h!]
\centering
\caption{Results of proposed approach.}
 \begin{tabular}{||c c c||} 
 \hline
 Word Embedding & Accuracy & F-measure \\ [0.5ex] 
 \hline\hline
 GLove & 89\%  & 86\% \\ 
\hline
 Word2vec & 87\% & 84\% \\
\hline
\end{tabular}
\label{table:1}
\end{table}

\begin{itemize}
    \item {This is the first approach that can accurately detect comprised accounts without a behavioral profile of the user. This will help in early detection very important in order to control the damage.}
    \item {This approach is useful for two categories of users: users with low profile accounts and users who do not have stable behavior.}
    \item{This approach can predict the authorship of posts that may not have been encountered within the training set.}
\end{itemize}

\section{Conclusion}
Today, social networks need an assistant system to detect compromised accounts, to prevent further damage caused by the accounts. The paper's key question is, in contexts of compromised accounts, whether it is possible to verify the authorship by examining the differences between the two posts. In this regard, a real-time approach using deep learning techniques is proposed to detect compromised accounts based on authorship verification of social media posts. The approach focuses on the last message that a user has posted on the social network. No additional information such as other posts,  behavioral profiles or social activity required. Model performance is evaluated by using a real dataset of Twitter compromised and not-compromised accounts. experiments on this dataset addressed The difference between the two posts can be distinguished by the features extracted using LSTM, Despite the short length of user-written posts. Also, the results showed that Glove representation generally performed better than Word2vec representations. For future works it would be of great interest to check the sensitivity of other languages to the user writing style to detect comprised accounts. 

\end{document}